\documentclass[%
 reprint,
 amsmath,amssymb,
 aps,
 prl,
 floatfix
]{revtex4-2}
\usepackage{physics}
\usepackage{graphicx}% Include figure files
\usepackage{dcolumn}% Align table columns on decimal point
\usepackage{bm}% bold math
\usepackage{color}
\usepackage[T1]{fontenc}

\usepackage{xr}
\makeatletter

\newcommand*{\addFileDependency}[1]{% argument=file name and extension
\typeout{(#1)}% latexmk will find this if $recorder=0
\@addtofilelist{#1}
\IfFileExists{#1}{}{\typeout{No file #1.}}
}\makeatother

\newcommand*{\myexternaldocument}[1]{%
\externaldocument[supp-]{#1}%
\addFileDependency{#1.tex}%
\addFileDependency{#1.aux}%
}
\myexternaldocument{supp}

\begin{document}

\title{Anomalous Thermal Transport of SrTiO$_3$ Driven by \\
Anharmonic Phonon Renormalization }% Force line breaks with \\

\author{Jian Han}
\affiliation{School of Materials Science and Engineering, Tsinghua University}
\affiliation{Graduate School of the China Academy of Engineering Physics, Beijing 100088, People’s Republic of China}
\author{Changpeng Lin}
\affiliation{Theory and Simulation of Materials (THEOS), and National Centre for Computational Design and Discovery of Novel Materials (MARVEL), \'Ecole Polytechnique F\'ed\'erale de Lausanne, 1015 Lausanne, Switzerland}
% \collaboration{MUSO Collaboration}%\noaffiliation
\author{Ce-wen Nan}
\affiliation{School of Materials Science and Engineering, Tsinghua University}
\author{Yuan-hua Lin}
\email{linyh@tsinghua.edu.cn}
\affiliation{School of Materials Science and Engineering, Tsinghua University}
\author{Ben Xu}
%  \homepage{http://www.Second.institution.edu/~Charlie.Author}
\email{bxu@gscaep.ac.cn}
\affiliation{
 Graduate School of the China Academy of Engineering Physics, Beijing 100088, People’s Republic of China
}%

\date{\today}% It is always \today, today,
\begin{abstract}
        SrTiO$_3$ has been extensively investigated owing to its abundant degrees of freedom for modulation.
        However, the microscopic mechanism of thermal transport especially the relationship between phonon scattering and lattice distortion during the phase transition are missing and unclear.
        Based on deep-potential molecular dynamics and self-consistent \textit{ab initio} lattice dynamics, we explore the lattice anharmonicity-induced tetragonal-to-cubic phase transition and explain this anomalous behavior during the phase transition.
        Our results indicate the significant role of the renormalization of 
        third-order interatomic force constants to second-order terms.
        Our work provides a robust framework for evaluating the thermal transport properties during structural transformation, benefitting the future design of promising thermal and phononic materials and devices.
\end{abstract}

\maketitle
%\tableofcontents

Perovskite materials have emerged as one of the most effective material in functional electronics ~\cite{yanLargeenergygapOxideTopological2013,panUltrahighEnergyDensity2019,zhengElectricalThermalTransport2021}.
However, their thermal conductivities during phase transition remain elusive because of the complex energy landscape and lattice anharmonicity. 
In particular, the complex patterns of oxygen octahedral distortions that introduce various structural phase transitions and consequently varying degrees of anharmonicity can lead to anomalous thermal transport behavior~\cite{tachibanaThermalConductivityPerovskite2008}.
For example, SrTiO$_3$ exhibits antiferrodistortion (AFD) in the low-temperature tetragonal phase, and as the temperature $T$ increases, the AFD amplitude and soft mode frequencies decrease to zero at approximately 105 K, resulting in a phase transition into a cubic phase~\cite{STO_Experiment_SoftModeAngle_01,STO_Experiment_SoftModeFrequency_01}.
The thermal conductivity $\kappa$ of SrTiO$_3$ exhibits a weak temperature dependence $\sim T^{-0.4}$ in the cubic phase and a continuous variation with a bump near the transition point~\cite{STO_Experiment_ThrmCond_02,STO_Experiment_ThrmCond_03}.
In contrast, lattice-anharmonicity-induced three-phonon interactions generally result in a $\sim T^{-1}$ trend in the temperature dependence of the thermal conductivity.
In previous studies, the weak temperature dependence was used to be explained by the degeneracy between the transverse optic and the longitudinal acoustic phonon branches\cite{STO_Experiment_ThrmCond_02},
nevertheless, this is based on an empirical observation and is not a complete picture.
To unveil the mechanism behind the anomalous behavior of $\kappa$ and its temperature dependence during phase transition, understanding the temperature dependence of lattice distortion and its influence on anharmonicity and thermal conductivity is important, which remains a challenging task.
\begin{figure*}[!htbp]
    \includegraphics{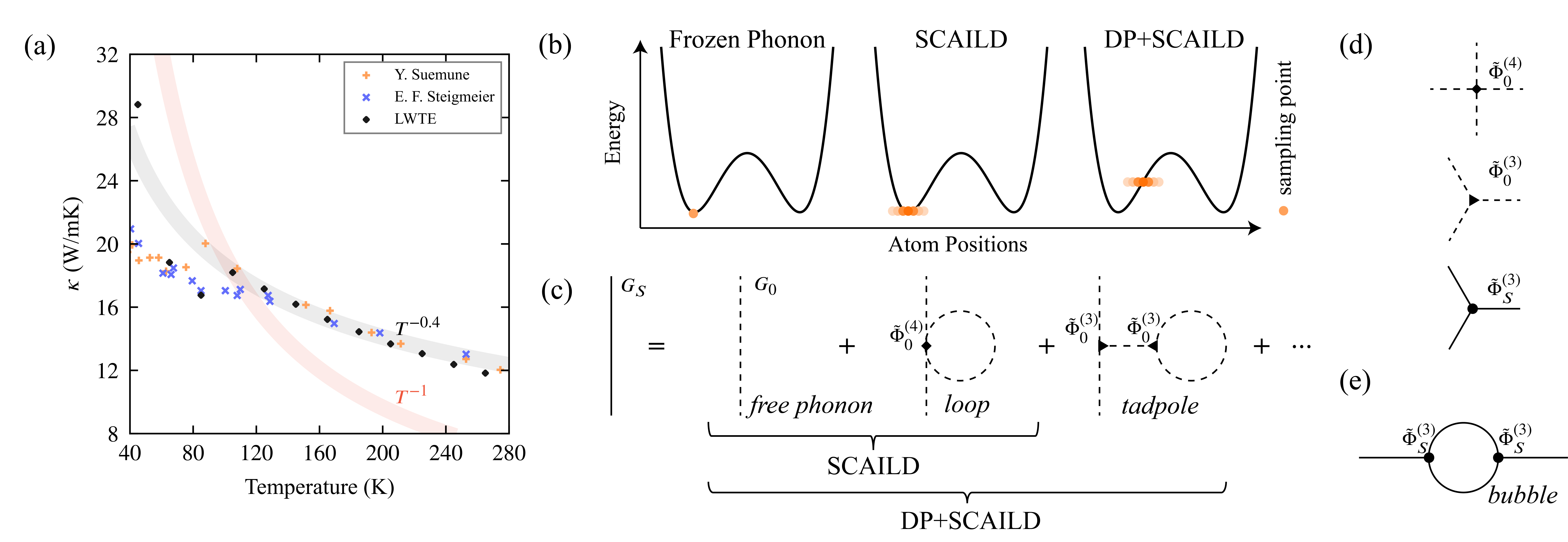}
    \caption{\label{fig_LWTE_vs_Experiments}
    (a) Comparison of calculated LWTE thermal conductivity with experimental results
    \cite{STO_Experiment_ThrmCond_02,STO_Experiment_ThrmCond_03}. 
    The LWTE thermal conductivity was calculated using 
    $\mathbf{r}(T+95\text{K})$, $\boldsymbol{\Phi}^{(2)}(T+95\text{K})$ and $\boldsymbol{\Phi}^{(3)}(T+95\text{K})$.
    (b) Schematic of frozen phonon method~\cite{ThrmCond_PhonoPy_01}, SCAILD, and our method combining DP and SCAILD.
    (c) Diagrammatic description of the relation between finite temperature phonon propagator $G_S$ calculated using SCAILD results and the free phonon propagator $G_0$.
    (d) Diagrammatic description of bare three-phonon and four-phonon vertex 
    and the effective three-phonon vertex at finite temperature.
    (e) Diagrammatic description of the bubble term to calculate the phonon lifetime in RTA.
    }
\end{figure*}
To accurately account for the temperature dependence of the thermal conductivity,
Bianco \textit{et al.}~\cite{SSCHA_bubble_term} provided a rigorous framework to accurately calculate the influence of anharmonicity and explain the details from the perspective of perturbation.
At the lowest perturbation level,
3\textsuperscript{rd}- and 4\textsuperscript{th}-order interatomic force constants (IFCs)
can introduce renormalizations through tadpole, loop, and bubble diagrams,
which result in the frequency shift and broadening of phonon quasiparticles.
Approaches based on many-body perturbation theory and frozen phonon method, such as ShengBTE\cite{ThrmCond_ShengBTE_01} and Phono3Py\cite{ThrmCond_phono3py_01},
consider only the broadening from the bubble term.
By contrast, the temperature-dependent effective potential (TDEP) method~\cite{TDEP_02,TDEP_03},
self-consistent \textit{ab initio} lattice dynamics (SCAILD)~\cite{SCAILD_01,SCAILD_02}
and self-consistent phonon (SCPH) theory~\cite{STO_Calc_Tadano_01} 
can only additionally consider the frequency shift from the loop diagrams, 
as the PES is usually expanded around the lattice structure at the ground state.
In order to take into account the contributions of the tadpole diagrams in those methods, 
one needs to also include the temperature dependence of lattice structure, 
which can be obtained from the NPT (constant particle number, pressure and temperature) ensemble averaging 
or free energy minimization~\cite{SSCHA_bubble_term}.
A sufficiently broad sampling is necessary in both methods for the lattice structure to converge.
However, such a sampling is extremely challenging, since predicting the phase transition process requires an accurate PES from \textit{ab initio} calculations.

In this study, we address this problem by integrating finite-temperature molecular dynamics (MD) simulations with SCAILD. 
To accurately describe the PES, we trained a deep learning model~\cite{DeePMD_01,DeePMD_02,GeTe_ML_ThrmCond_01} based on \textit{ab initio} calculations for SrTiO$_3$.
The lattice structure at a certain temperature was determined by MD with a quantum thermal bath (QTB)~\cite{MD_lammps_01,MD_lammps_QTB_Dammak_01,MD_lammps_QTB_Barrat_02}.
Thereafter, we fitted the effective IFCs via the SCAILD method and obtained their temperature dependence.
We solved the linearized Wigner transport equation (LWTE) to calculate the lattice thermal conductivity.
Our results agree well with the experimental measurements among a wide temperature range near the phase transition\cite{STO_Experiment_ThrmCond_02,STO_Experiment_ThrmCond_03}.
To identify the origin of the thermal conductivity anomaly, we investigated the importance of the effective 2\textsuperscript{nd}- and 3\textsuperscript{rd}- IFCs, group velocities, and relaxation times.
To demonstrate the importance of the temperature dependence of the lattice structure, 
we also compared the effective 2\textsuperscript{nd}-order IFCs with and without the temperature dependence of the lattice structure.
Our methodology offers a systematic approach for studying the thermal conductivity and lattice dynamics-related properties during the phase-transition process.
We studied the temperature dependence of the lattice structure 
by considering the NPT ensemble average of the atomic positions and volume in MD simulations as implemented LAMMPS~\cite{MD_lammps_01} package.
Moreover, we incorporated the quantum effect using a quantum thermal bath~\cite{MD_lammps_QTB_Dammak_01,MD_lammps_QTB_Barrat_02,wuLargescaleAtomisticSimulation2022} method. 
To describe the interatomic potential energy surface (PES), 
we used a deep neural network (DNN)-based 
Deep Potential (DP) model~\cite{DeePMD_01,DeePMD_02}. 
This is consistent with the first-principles results with only 1.21 meV energy error per atom. 
This model was trained for $1\times 10^{6}$ steps, with 
more than 2000 labeled data points generated iteratively 
using a Deep Potential GENerator (DP-GEN)~\cite{DeePMD_DPGEN_01,DeePMD_DPGEN_02} platform.
The data were labeled via DFT methods~\cite{DFT_01} from a first-principles perspective.
Vienna \textit{Ab initio} Simulation Package (VASP)~\cite{DFT_vasp_01,DFT_vasp_02,DFT_vasp_03,DFT_vasp_04}
was used to perform DFT calculations via
the projector-augmented wave (PAW) method~\cite{DFT_vasp_PAW}.
A revised version of the Perdew-Burke-Ernzerhof~\cite{DFT_vasp_pbe_01,DFT_vasp_pbesol_01} paraterization of the generalized gradient approximation~\cite{DFT_vasp_gga_01} was applied to describe the exchange-correlation effects of electrons. 
In all DFT calculations, the plane-wave kinetic energy cutoff was set to 700 eV, 
and k-spacing was less than 0.14 \AA$^{-1}$.
The temperature dependence of the lattice structure shows that
the phase transition temperature in the MD simulation is approximately 200 K,
which is higher than the experimental result of 105 K.
Such overestimation
could be attributed to the limited precision of the PBEsol pseudopotential~\cite{tadanoFirstPrinciplesLatticeDynamics2018}
or the zero-point energy leakage in QTB~\cite{brieucZeroPointEnergyLeakage2016}.
Apart from this slight discrepancy, our results capture the phase-transition process
and enable us to further delve into the origin of the anomalous thermal conductivity in SrTiO$_3$.

\begin{figure*}[!htbp]
    \includegraphics{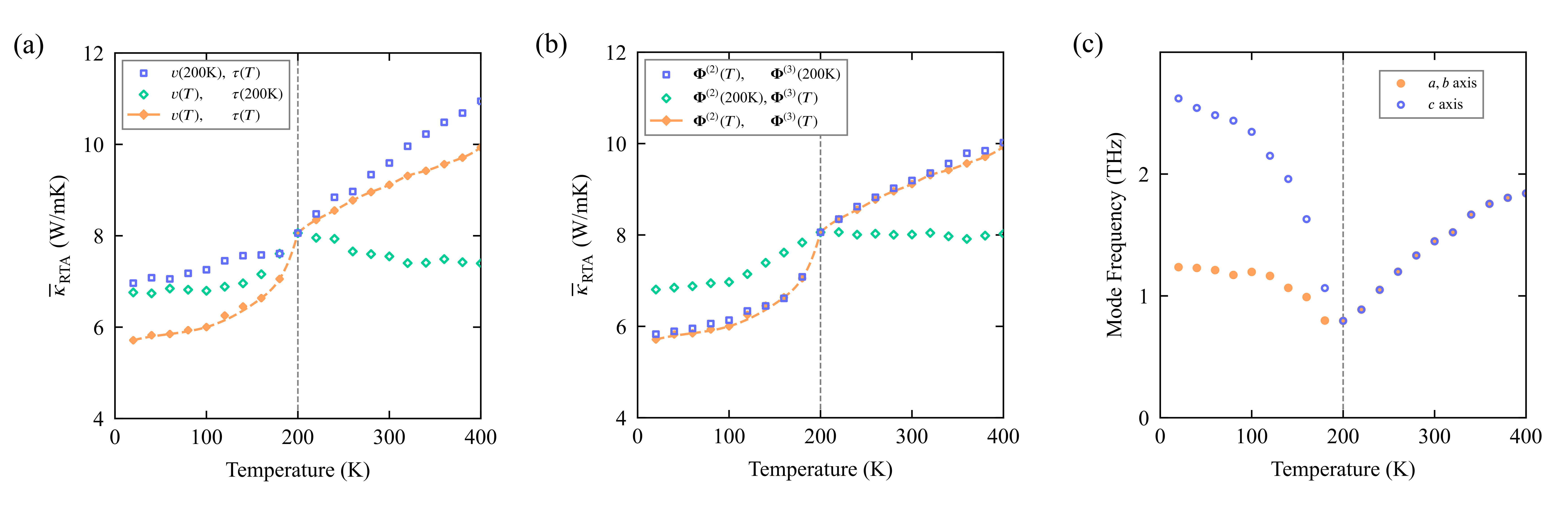}
    \caption{\label{fig_soft_mode_frequency_and_LBTE_scaled_kappa_vs_T} 
    (a) The temperature dependence 
    of scaled thermal conductivity $\overline{\kappa}$.
    For orange diamonds\, the temperature dependence of $v$ and $\tau$ are considered..
    For blue cubic, to show the impact of the temperature dependence of $v$,
    $\tau$ are kept the same.
    For green diamonds, to show the impact of the temperature dependence of $\tau$ and $v$ are kept the same.
    (b) The temperature dependence 
    of scaled thermal conductivity $\overline{\kappa}$.
    For orange diamonds, the temperature dependence of $\mathbf{r}$, $\boldsymbol{\Phi}^{(3)}$ and $\boldsymbol{\Phi}^{(2)}$ are considered.
    For blue cubic, $\mathbf{r}$ and $\boldsymbol{\Phi}^{(3)}$ are fixed to show the impact of the temperature dependence of $\boldsymbol{\Phi}^{(2)}$.
    For green diamonds, $\mathbf{r}$ and $\boldsymbol{\Phi}^{(2)}$ are fixed to show the impact of the temperature dependence of $\boldsymbol{\Phi}^{(3)}$.
    (c) The temperature dependence of AFD mode frequencies.
    }
\end{figure*}

To investigate the temperature dependency of the 2\textsuperscript{nd}- and 3\textsuperscript{rd}-order IFCs, 
we applied the SCAILD method to fit them iteratively 
with a temperature-dependent lattice structure $\mathbf{r}$ of SrTiO$_3$.
Sampling structures were randomly generated with the following probability:
$\rho(\{ u^{a} \}) \propto  \exp (-\frac{1}{2} \mathbf{u} \boldsymbol{\Psi}^{-1} \mathbf{u})$, where the quantum covariance is defined as
\begin{eqnarray}
    \begin{aligned}
        \label{eq_SCAILD_covariance_matrix}
        \Psi^{ab}  & = 
        \sum_{\nu}
        \frac{\hbar}{2\sqrt{m_{a} m_{b}}} 
        \frac{ 1 + 2 n_\nu }{\omega_{\textbf{} \mathbf{}\nu} } 
        \varepsilon^{ a} \phantom{}_{\nu}
        \varepsilon^{b}  \phantom{}_{\nu}.
    \end{aligned}
\end{eqnarray}
In the above equation, $u^{a}$ is the displacement of atom $i$ in $\alpha$ direction relative to the finite-temperature equilibrium position $r^{a}$, 
$a$ is the shorthand of $i\alpha$,
$m_a$ is the atomic mass, 
$\nu$ is the shorthand of $\mathbf{q}\lambda$,
$\varepsilon^{a} \phantom{}_{\nu}$ is the wavefunction of $\lambda$-branch phonon on wavevector $\mathbf{q}$, and
$n_\nu$ is the Bose-Einstein distribution.
The interatomic forces of the random structures were calculated using the DP model. 
We fitted the 2\textsuperscript{nd}- and 3\textsuperscript{rd}-order IFCs at the given temperature $T$ with a cutoff radii of 7.0 {\AA} and 3.0 {\AA}, respectively, denoted as $\boldsymbol{\Phi}^{(2)}$ and $\boldsymbol{\Phi}^{(3)}$, via least-squares minimization, 
with the constraints of space-group symmetry, permutation and acoustic sum rules~\cite{CSLD_01,CSLD_02}.
Thereafter, we calculated the thermal conductivity tensor $\kappa^{\alpha\beta}$ as a function of temperature by solving the linearized Wigner equation (LWTE) using the Phono3Py package~\cite{ThrmCond_phono3py_01} with a $\mathbf{q}$-mesh of $13 \times 13 \times 13$.
The thermal conductivity was subsequently calculated by considering the average of the diagonal part of the thermal conductivity tensor.
To compare our results with the experimental measurements, 
we shifted the phase transition point.
For example, the thermal conductivity at 105~K can be evaluated using 
$\mathbf{r}$, $\boldsymbol{\Phi}^{(2)}$, and $\boldsymbol{\Phi}^{(3)}$ at 200~K and the Bose-Einstein distribution at 105~K.
In Fig.~\ref{fig_LWTE_vs_Experiments}(a), we compare the simulated thermal conductivity with previous experimental  measurements~\cite{STO_Experiment_ThrmCond_01,STO_Experiment_ThrmCond_02,STO_Experiment_ThrmCond_03}.
In our calculations, we capture exactly two features.
First, the thermal conductivity demonstrates continuous variation and exhibits a bump at the phase transition point.
Second, the temperature dependence in the high-temperature cubic phase agrees well with the experiment and exhibits a tendency of $\sim T^{-0.4}$.
Through calculations, we can control the variables to better understand the causes of thermal conductivity anomalies.
In the following calculations, we used the relaxation-time approximation (RTA)
and calculate the thermal conductivity as follows:
\begin{eqnarray}
    \label{eq_LBTE_RTA}
    \begin{aligned}
    \kappa^{\alpha\beta}_{\text{RTA}}
        = 
        \frac{1}{N_0 \Omega} \sum_{\nu}
            C_\nu
                v^{\alpha\nu}
                    v^{\beta\nu}
                        \tau_\nu,
    \end{aligned}
\end{eqnarray}
where $v^{\alpha\nu}$ denotes the phonon group velocity,
$C_v$ is the phonon heat capacity, $\tau_{\nu}$ is the phonon relaxation time, and
$N_0$ is the total number of $\mathbf{q}-$point in reciprocal space.
The relaxation time $\tau_{\nu} = \left[2\Gamma_\nu(\omega_\nu)\right]^{-1}$ 
can be calculated using the imaginary part $\Gamma_\nu(\omega)$ of the bubble’s self-energy ${\Sigma}^{\text{bubble}}_{\nu}(\omega)$
as shown in Fig.~\ref{fig_LWTE_vs_Experiments}(e).
The bubble self-energy is expressed as follows:

\begin{eqnarray}
   \begin{aligned}
       \label{eq_bubble_expression} 
       {\Sigma}^{\text{bubble}}_{\nu}(\omega)= 
         & \frac{1}{2 \hbar} 
           \sum_{\substack{\nu_{1}, \nu_{2}, s=\pm 1 \\ \mathbf{q}_1+\mathbf{q}_2+\mathbf{q}=0}}
         \left|\tilde{\Phi}^{(3)}_S \left(-\nu ; \nu_1 ; \nu_2 \right)\right|^{2} \\
          \times&\left[
           % \frac{\left(n_{\nu_1}+n_{\nu_2}+1\right)}{s (\omega_{\nu}+i0^{+})+\omega_{\nu_{1}}+\omega_{\nu_{2}}}
         % - \frac{\left(n_{\nu_1}-n_{\nu_2}\right)}{s (\omega_{\nu}+i0^{+})+\omega_{\nu_{1}}-\omega_{\nu_{2}}}\right]
         \frac{\left(n_{\nu_1}+n_{\nu_2}+1\right)}{s \omega^\prime+\omega_{\nu_{1}}+\omega_{\nu_{2}}}
         - \frac{\left(n_{\nu_1}-n_{\nu_2}\right)}{s \omega^\prime+\omega_{\nu_{1}}-\omega_{\nu_{2}}}\right],
   \end{aligned}
\end{eqnarray}
where $\tilde{\boldsymbol{\Phi}}^{(3)}_S$ denotes the finite-temperature three-phonon vertex (see Eq.~\eqref{supp-eq_ph_ph_interaction_strength}), and 
$\left|\tilde{\boldsymbol{\Phi}}^{(3)}_S\right|^2$represents the 
three phonon-phonon (\textit{ph-ph}) interaction strength.
The subscript $S$ denotes the finite-temperature results from SCAILD, and $\omega^\prime = \omega + i0^{+}$, where $0^{+}$ is a positive infinitesimal.
To identify the role of temperature dependence of various physical quantities, 
we defined the scaled thermal conductivity $\overline{\kappa}$, 
where the Bose-Einstein distribution $n_\nu$ is fixed at 300 K.
As can be seen in Fig.~\ref{fig_soft_mode_frequency_and_LBTE_scaled_kappa_vs_T}(a) and (b), the orange diamond represents the scaled thermal conductivity with the temperature dependence of other parameters considered explicitly.
Such a scaled thermal conductivity can directly demonstrate the influence from the phonon group velocities and lifetimes, without 
the normal $\sim T^{-1}$ dependence owing to an increase in $n_\nu$.
Therefore, the temperature dependence of $\overline{\kappa}$ 
can recognize the reason for this anomaly in thermal conductivity of SrTiO$_3$.

To extract the respective contributions to thermal conductivity, we fixed one of the parameters in turn.
We find that only at fixed phonon lifetimes, 
(green diamonds in Fig.~\ref{fig_soft_mode_frequency_and_LBTE_scaled_kappa_vs_T}(a)),
$\overline{\kappa}$ increases and subsequently decreases with an increase in temperature 
because of the change in group velocities. 
Thus, we determine that bump is contributed by the phonon velocity.
Additionally, we find that at the phase transition point, higher group velocities  
result from a lower AFD soft-mode frequency, as shown in 
Fig.~\ref{fig_soft_mode_frequency_and_LBTE_scaled_kappa_vs_T}(c).
The AFD mode frequencies are overestimated in our results; that is, they did not reach zero at the phase transition pint.
Such prediction of soft-mode frequency is limited by the computational cost of the phonon shift owing to different order bubble terms~\cite{SSCHA_bubble_term}.
At fixed phonon group velocities,
(the blue cubes in Fig.~\ref{fig_soft_mode_frequency_and_LBTE_scaled_kappa_vs_T}(a)), 
$\overline{\kappa}$ increases with the temperature, 
and this part of the increase 
counteracts a part of the normal $\sim T^{-1}$ dependence of $\kappa$.
By comparing $\overline{\kappa}$ with constant $\textit{ph-ph}$ interaction strength
as shown in Fig.~\ref{supp-fig_RTA_const_pp}, 
we also find that its the decreasing $\textit{ph-ph}$ interaction strength contributes to the raise of $\overline{\kappa}$.
Accordingly, we confirm that the decreasing $\textit{ph-ph}$ interaction strength contributes to a weak temperature dependence of $\kappa$ in the cubic phase, which is in contrast to previous understanding that
it can be explained by the degeneracy.

\begin{figure}[tbp]
    \includegraphics{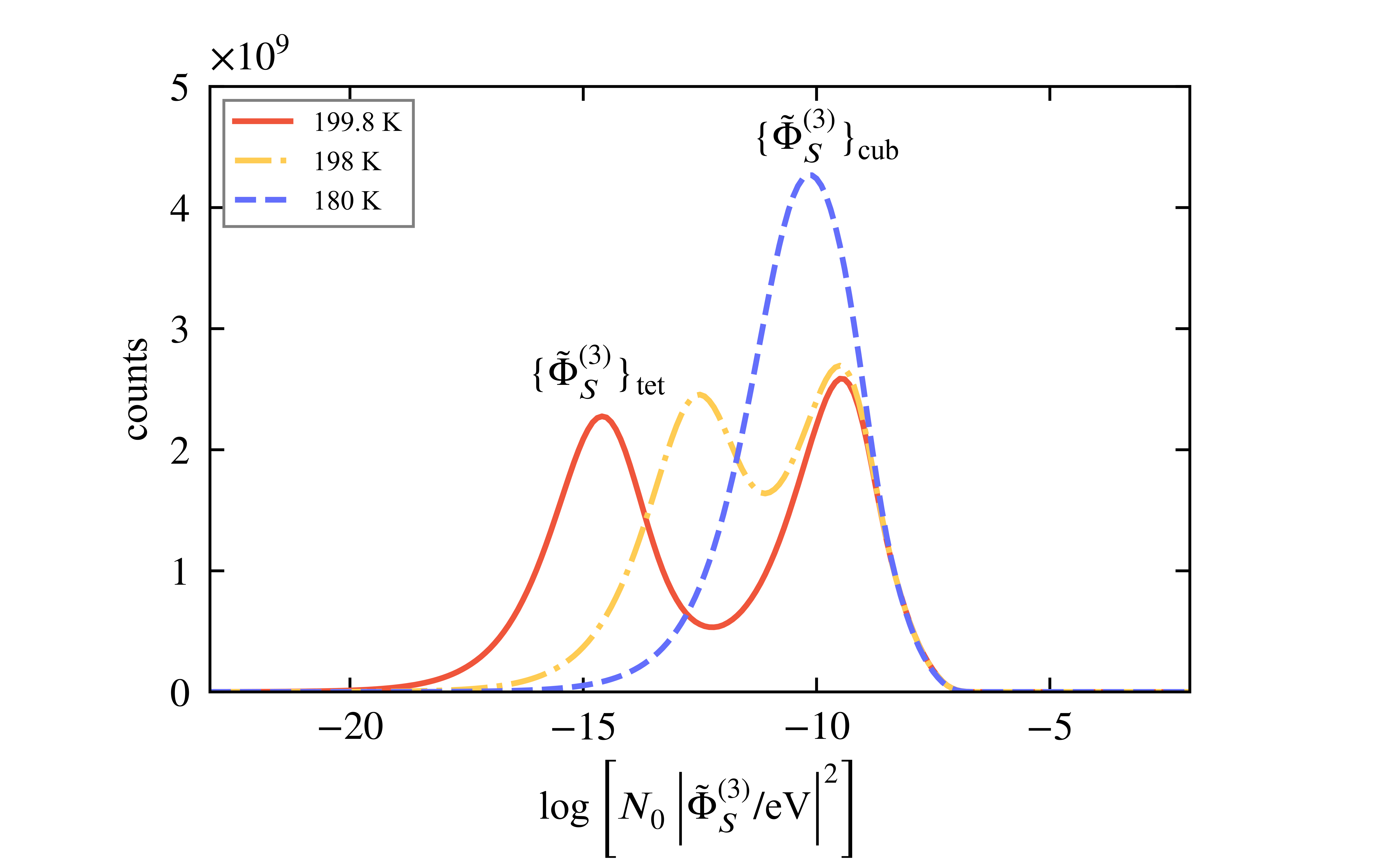}
    \caption{\label{fig_pp} 
    The x-axis is the \textit{ph-ph} interaction strengths, 
    and the y-axis shows the counts of each value.
    $\tilde{\boldsymbol{\Phi}}^{(3)}_S$ is calculated using $\mathbf{r}$, $\boldsymbol{\Phi}^{(2)}$ and $\boldsymbol{\Phi}^{(3)}$.
    The IFCs in 198 K and 199.8 K are interpolated from 180 K and 200 K results.
    }
\end{figure}

The $\textit{ph-ph}$ interaction strength is mainly determined 
by the effective 2\textsuperscript{nd}- and 3\textsuperscript{rd}-order IFCs.
To follow the temperature dependence of the 2\textsuperscript{nd}- and 3\textsuperscript{rd}-order effective IFCs 
on $\overline{\kappa}$, we performed the following procedure:
First, we fixed the 2\textsuperscript{nd}-order IFCs 
and maintained the temperature dependence of 3\textsuperscript{rd}-order IFCs.
The corresponding $\overline{\kappa}$ is 
shown in Fig.~\ref{fig_soft_mode_frequency_and_LBTE_scaled_kappa_vs_T}(b) using green diamonds.
We observe an increasing trend of $\overline{\kappa}$ only in the tetragonal phase 
but not in the cubic phase.
This implies that the weak temperature dependence of $\kappa$ in the cubic phase
is not due to the renormalization to the 3\textsuperscript{rd}-order IFCs.
Accordingly, we fix the 3\textsuperscript{rd}-order effective IFCs and maintain the temperature dependence of 2\textsuperscript{nd}-order IFCs.
The results are shown in Fig.~\ref{fig_soft_mode_frequency_and_LBTE_scaled_kappa_vs_T}(b) 
by the blue cubic.
Similar tendencies to orange diamonds
show that the anomalies in $\kappa$,
including bump and weak temperature dependences
are mainly caused by the renormalization to the 2\textsuperscript{nd}-order IFCs. 
Therefore, we investigated the \textit{ph-ph} interaction strength to clarify the reason behind the continuous change in the thermal conductivity at the phase transition point.
Fig.~\ref{fig_pp} shows the number of scattering channels with different interaction strengths at different temperatures.
Here, $\{\tilde{\Phi}^{(3)}_S\}_\text{tet}$ is the set of \textit{ph-ph} interaction vertices of scattering channels activated only in the tetragonal phase, and $\{\tilde{\Phi}^{(3)}_S\}_\text{cub}$ is one of the activated scattering channels in the cubic phase.
At 180 K, the peaks of $\{\tilde{\Phi}^{(3)}_S\}_\text{cub}$ and $\{\tilde{\Phi}^{(3)}_S\}_\text{tet}$ overlap, 
and the average value of $\{\tilde{\Phi}^{(3)}_S\}_\text{tet}$ is smaller than that of $\{\tilde{\Phi}^{(3)}_S\}_\text{cub}$.
With an increase in the temperature, the two peaks separate, and the average value of $\{\tilde{\Phi}^{(3)}_S\}_\text{tet}$ gradually approaches zero.
Because $\{\tilde{\Phi}^{(3)}_S\}_\text{tet}$ are already zero at the phase-transition point,
the sudden change in the scattering channel numbers owing to translational symmetry breaking
does not cause a sudden change in the phonon relaxation time, 
causing the thermal conductivity at the phase transition to change continuously.
According to Eq.~\eqref{supp-eq_ph_ph_interaction_strength}, the continuous change in $\{\tilde{\Phi}^{(3)}_S\}_\text{tet}$ to zero could happen only with 
a continuous decrease in the anisotropy of effective IFCs,
especially the 2\textsuperscript{nd}-order ones.
We find that to capture this behaviour, it's necessary to introduce the temperature dependence of the lattice structure.
At the lowest perturbation limit, 
the effective 2\textsuperscript{nd}-order IFCs expanded around
temperature-dependent lattice structure $\mathbf{r}$ are
\begin{eqnarray}
\label{eq_finite_temperature_IFCs_in_real_space}
\Phi^{(2)}_{a b}=\Phi^{(2)}_{0,a b} +\sum_{c} \Phi_{0,a b c}^{(3)} \delta r^{c} 
+\frac{1}{2} \sum_{c d} {\Phi}^{(4)}_{0,a b c d} \Psi_0^{c d},
\end{eqnarray}
where $\boldsymbol{\Phi}^{(n)}_0$ is the bare $n^{\text{th}}$-order IFCs expanded around atomic position
$\mathbf{r}_0$ at the ground state,
$\Psi_0^{c d}$ is the temperature-dependent quantum covariance calculated using $\boldsymbol{\Phi}^{(2)}_0$ according to 
Eq.~\eqref{eq_SCAILD_covariance_matrix} and
$\delta \mathbf{r}$ is the finite temperature atomic displacement. 
At the lowest perturbation limit, $\delta \mathbf{r}$ is\cite{SSCHA_bubble_term}:
\begin{eqnarray}
\delta r^{c}&=& r^{c} - r_0^{c} \nonumber \\
&=&-\frac{1}{2} \sum_{a b d} \mathrm{inv}\left[\Phi^{(2)}_0\right]{} ^{c a}
\Phi^{(3)}_{0,a b d} \Psi_0^{b d},
\end{eqnarray}
where $\mathrm{inv}\left[\Phi^{(2)}_0\right]{} ^{ca}$ is the $ca{}^\text{th}$-component of the inverse of $\boldsymbol{\Phi}^{(2)}_0$.

This renormalization to the $2^{\text{nd}}$-order IFCs can therefore be expressed using diagrammatic description, by introducing the tadpole and loop self-energy, denoted as ${\Sigma}^{\text{tadpole,0}}_{\nu}$ and 
${\Sigma}^{\text{loop,0}}_{\nu}$, to phonon quasiparticle, as shown in Fig.~\ref{fig_LWTE_vs_Experiments}(c). 
The corresponding finite-temperature phonon propagator is now $ G_{S} (\omega) \approx  G_{0} (\omega) +  G_{S} (\omega) =  G_{0} (\omega) + G_{0} (\omega) \left[ {\Sigma}^{\text{loop,0}}_{\nu} +{\Sigma}^{\text{tadpole,0}}_{\nu}  \right] G_{0} (\omega)$, where $G_{0}(\omega)$ denotes a free phonon propagator.
%determined by $\boldsymbol{\Phi}^{(2)}$ is renormalized by ,

%
\begin{figure}[t]
    \includegraphics{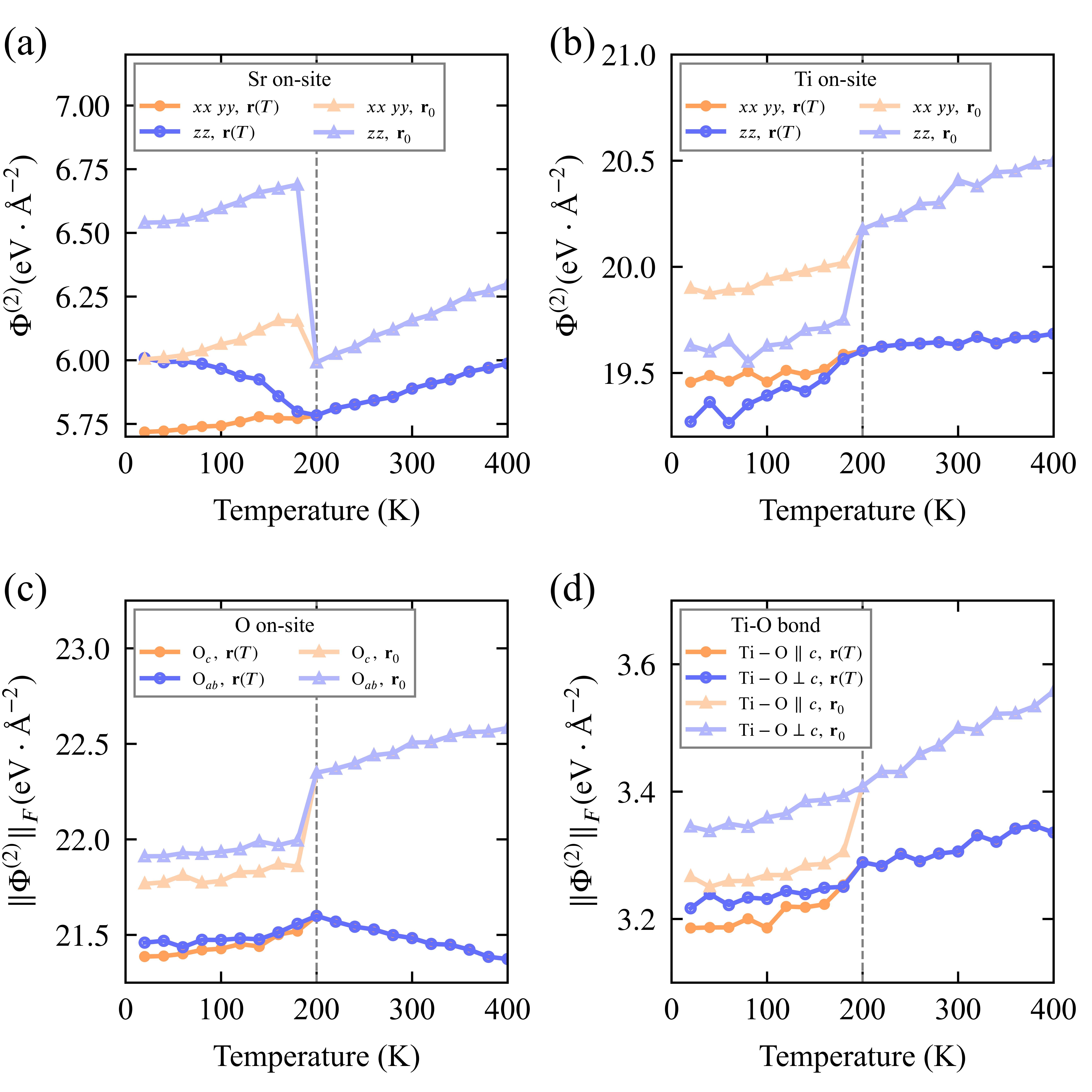}
    \caption{\label{fig_fc2_vs_T}
    Temperature dependence of 2\textsuperscript{nd}-order IFCs $\boldsymbol{\Phi}^{(2)}$.
    The dark and light symbols are for $\boldsymbol{\Phi}^{(2)}$ expanded 
    around $\mathbf{r}$ and $\mathbf{r}_0$, respectively.
    $\mathbf{r}_0$ was set to the lattice structure at the ground state of the tetragonal phase for $T<$200~K, and one of the cubic phase for $T\geq$200~K.
    We appoint the long axis of tetragonal phase $c$ along $z$ direction.
    (a) and (b) show the on-site terms of $\boldsymbol{\Phi}^{(2)}$ for Sr and Ti atom, respectively.
    Orange symbols indicate the $xx$ and $yy$ component, while the blue symbol indicates the $zz$ component.
    (c) and (d) show the Frobenius norm of $\boldsymbol{\Phi}^{(2)}$ for the on-site term of O atom 
    and nearest Ti-O bond, respectively.
    Orange symbols show the results of O atom whose nearest Ti-O bond is along the $c$ direction.
    Blue symbols show the results of O atom whose nearest Ti-O bond is along the $a$ or $b$ direction.  
    }
\end{figure}
Beyond the lowest perturbation limit, 
higher-order tadpole and loop diagrams also contribute to effective IFCs.
We include the contributions above by expanding the PES around the temperature-dependent lattice structure $\mathbf{r}$ using SCAILD, corresponding to the dark blue and orange circles in Fig.~\ref{fig_fc2_vs_T}.
In comparison, if the PES were expanded around $\mathbf{r}_0$,
the contribution of the tadpole term is neglected, while the contributions from the loop terms are still maintained, which corresponds to the light blue and orange triangles in Fig.~\ref{fig_fc2_vs_T}.
Comparing these two sets of curves, a sharp change in $\boldsymbol{\Phi}^{(2)}$ near the phase transition point is found when the tadpole diagrams are neglected, 
while $\boldsymbol{\Phi}^{(2)}$ expanded around $\mathbf{r}$ gradually approached the value in the cubic phase with a decreasing degree of anisotropy.
We conclude that the tadpole diagram term 
determines the continuous change in the thermal conductivity at the phase transition point.
In summary, we have calculated the thermal conductivity of SrTiO$_3$
via the SCAILD method with a temperature-dependent lattice structure obtained from DPMD.
We explain the origin of the thermal conductivity anomalies in SrTiO$_3$.
The continuous variation with a bump at the phase transition point is mainly driven by the temperature dependence of the group velocities,
whereas the weak temperature-dependent results come from the decreasing \textit{ph-ph} interaction strength,
which is in contrast to the previous understanding that it was explained by the degeneracy of phonon modes.
Both phenomena are caused by the renormalization to the effective 2\textsuperscript{nd}-order IFCs by anharmonicity, where the tadpole self-energy term contributed by 3\textsuperscript{rd}-order IFCs cannot be neglected. 
%
%Our study has certain limitations, such as the overestimation of phase transition point and AFD mode frequencies, that is, the AFD mode frequencies did not reach zero in our results.
%
%The current prediction of soft-mode frequency is limited by the computational cost of the phonon shift owing to different order bubble terms~\cite{SSCHA_bubble_term}.
%
Our study provides a complete prediction and systematic analysis of the anomalous thermal conductivity of SrTiO$_3$.
The phonon renormalization can also have a significant impact on other transport properties, such as the
electrical transport limited by electron-phonon interactions.
Therefore, we believe that computational framework adopted in this work promises tremendous potential for predicting the transport behavior of future functional materials.

This work was funded by the National Natural Science Foundation of China (Grant Nos. 52072209, 51790494 and 12088101).
%

% \nocite{*}
% \clearpage
\bibliographystyle{apsrev4-2}
\bibliography{main}% Produces the bibliography via BibTeX.
\end{document}

% \title{Harmonic Eigenvector Driven Thermal Conductivity Deviation from 1/T Rule}% Force line breaks with \\
\title{\normalfont{Supplementary Information for} Anomalous Thermal Transport of SrTiO$_3$ Driven by Anharmonic Phonon Renormalization}% Force line breaks with \\

\author{Jian Han}
\affiliation{School of Materials Science and Engineering, Tsinghua University}
\author{Changpeng Lin}
\affiliation{École Polytechnique Fédérale de Lausanne}
\author{Ben Xu}
\email{bxu@gscaep.ac.cn}
\affiliation{Graduate School of China Academy of Engineering Physics}
\author{Yuanhua Lin}
\email{linyh@tsinghua.edu.cn}
\affiliation{School of Materials Science and Engineering, Tsinghua University}
\date{\today}% It is always \today, today,

%
\renewcommand{\thefigure}{S\arabic{figure}}
\renewcommand{\theequation}{S\arabic{equation}}
\renewcommand \thesection{S\arabic{section}}

\maketitle
%\tableofcontents
\section{Accuracy of the Deep Potential Model}
\begin{figure}[!hb]
        \includegraphics{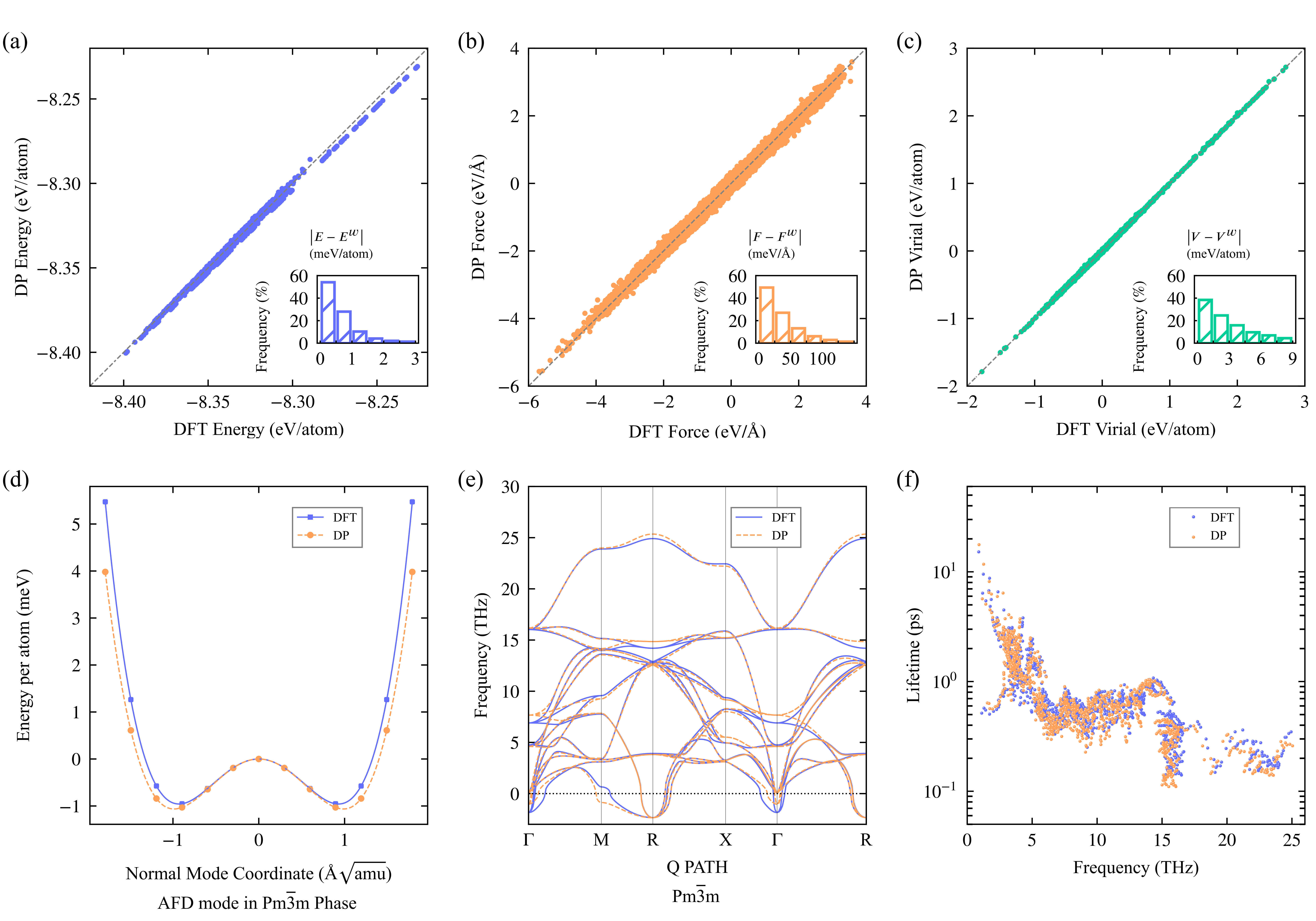}
        \caption{\label{fig_DP_results}
        Comparison of results calculated using the deep potential (DP) model and density functional theory (DFT). (a) energies, (b) atomic forces, (c) viral for all configurations in the training and testing dataset, (d) energy double well of antiferrodistortion (AFD) mode, (e) phonon spectrum in cubic phase using frozen phonon method~\cite{ThrmCond_PhonoPy_01} and (f) phonon lifetime in cubic phase using self-consistent \textit{ab initio} lattice dynamics (SCAILD) method~\cite{SCAILD_01,SCAILD_02} and Phono3Py package~\cite{ThrmCond_phono3py_01}.
        }
    \end{figure}

    The DP model model was trained for $1\times 10^{6}$ steps, with 
    more than 2000 labeled data points generated iteratively 
    using a Deep Potential GENerator (DP-GEN)~\cite{DeePMD_DPGEN_01,DeePMD_DPGEN_02} platform.
    The cutoff radius is 9~{\AA} and the smoothing starts at 2~{\AA}.
    The initial learning rate was 0.001 and the final learning rate was less than $4\times10^{-8}$.
    This DP model is consistent with the first-principles results with only 1.21 meV energy error per atom.
    To show the accuracy of our model, we compare the calculation results using the DP model and DFT method in Fig.~\ref{fig_DP_results}.

\section{Temperature Dependent Lattice Structure Using Molecular Dynamics Simulation}
    \begin{figure}[!hb]
        \includegraphics{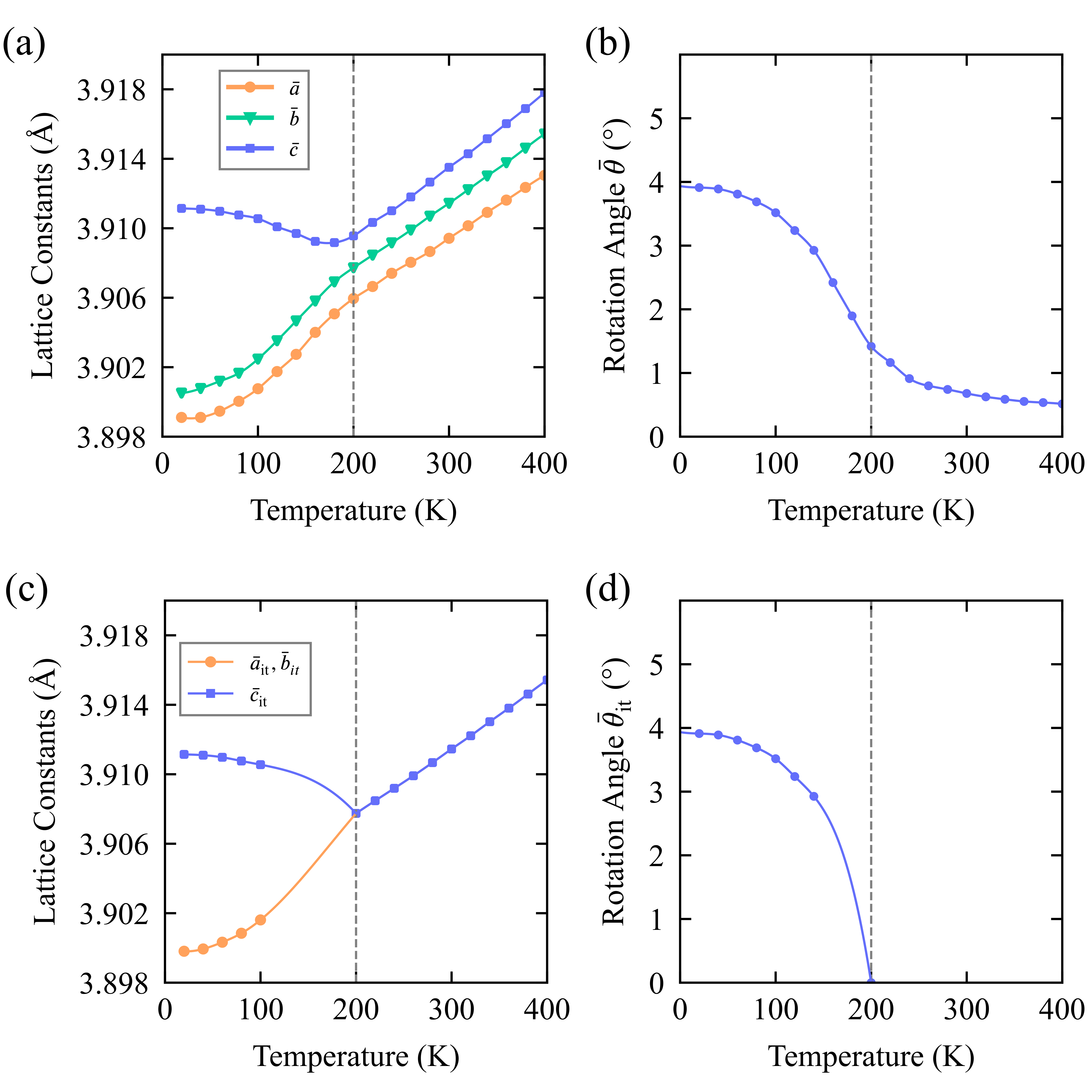}
        \caption{\label{fig_MD_results_vs_T}
        (a)(b) The temperature dependence of average ordered lattice constants and absolute value of tilting angle in the DP model based MD simulations under QTB.
        (c)(d) The temperature dependence of interpolated lattice constants and tilting angle.
        }
    \end{figure}

    NPT molecular dynamics (MD) simulations were performed with periodic boundary conditions 
    and at zero pressure using LAMMPS package~\cite{MD_lammps_01} 
    with a quantum thermal bath (QTB)~\cite{MD_lammps_QTB_Dammak_01,MD_lammps_QTB_Barrat_02}.
    The time step was set to 1~fs and the size of supercell was set to $10\times10\times10$ (5000 atoms).
    The temperatures were set from 20~K to 400~K with 20~K step.
    The system reached equilibrium in the first 10~ps.
    Then the lattice structures were output every 80~fs in the following 160~ps.

    Because we observe a long axe switching behavior in the MD simulation of SrTiO$_3$ system based on deep potential (DP) model, which also appeared in previous studies
    \cite{STO_Calc_HeRi_01},
    we cannot average the atomic positions directly.
    Therefore we defined the ordered lattice constant $\overline{a}$, $\overline{b}$ and $\overline{c}$, 
    and absolute value of tilting angle $\overline{\theta}$ of oxygen octahedral 
    at each temperature as follows:
    \begin{eqnarray}
    \begin{aligned}
        \label{eq_STO_MD_afterprocess}
        \left[\overline{a}(T),\overline{b}(T),\overline{c}(T)\right]
            =
            \frac{\Delta t}{N_t}\sum_t \textrm{Sort}[a(T,t),b(T,t),c(T,t)]\\
        \overline{\theta}(T)
            =
            \frac{\Delta t}{N_t}\sum_t \max\left[\left|\theta_a(T,t)\right|,\left|\theta_b(T,t)\right|,\left|\theta_c(T,t)\right|\right],
    \end{aligned}
    \end{eqnarray}
    where $\theta_a,\theta_b$ and $\theta_c$ are tilting angles around corresponding axis,
    $a,b$ and $c$ are lattice constants outputted during MD simulations, 
    $T$ is the temperature and $t$ is the simulation time. 
    Fig.~\ref{fig_MD_results_vs_T}(a)(b) shows the temperature dependence of $\overline{a},\overline{b},\overline{c}$ and $\overline{\theta}$, the phase transition temperature in MD simulation is around 200 K. 
    The interpolation was then performed 
    to get the lattice constant and tilting angle at each temperature, 
    which would be used in the further SCAILD calculations.
    The results were shown in Fig.~\ref{fig_MD_results_vs_T}(c)(d).
    The data used in interpolation are 
    $\overline{a}_\text{it}$, $\overline{b}_\text{it}$, $\overline{c}_\text{it}$ and $\overline{\theta}_\text{it}$.
    We choose $\overline{a}_\text{it}$, $\overline{b}_\text{it}$, $\overline{c}_\text{it}$ and $\overline{\theta}_\text{it}$ as follows:
    \begin{eqnarray}
    \begin{aligned}
    \begin{cases}
        \overline{\theta}_\text{it}(T) = \overline{\theta}(T), &T\leq 140\ K \\
        \overline{\theta}_\text{it}(T) = 0,                    &T\geq 200\ K \\
    \end{cases}
    \end{aligned}
    \end{eqnarray}
    \begin{eqnarray}
    \begin{aligned}
    \begin{cases}
        \overline{a}_\text{it}(T) = \overline{b}_\text{it}(T) = \sqrt{\overline{a}(T)\overline{b}(T)}, &T\leq 100\ K \\
        \overline{a}_\text{it}(T) = \overline{b}_\text{it}(T) = \overline{c}_\text{it}(T) = \sqrt[3]{\overline{a}(T)\overline{b}(T)\overline{c}(T)}.                    &T\geq 200\ K \\
    \end{cases}
    \end{aligned}
    \end{eqnarray}

\section{Details in Calculating Scaled Thermal Conductivity}

    Under relaxation time approximation, the lattice thermal conductivity using linearized Boltzmann transport equation is:
    \begin{eqnarray}
    \label{eq_LBTE_RTA}
    \begin{aligned}
    \kappa^{\alpha\alpha'}_{\text{RTA}}
        = 
        \frac{1}{N_0 \Omega} \sum_{\nu}
            C_\nu
                v^{\alpha,\nu}
                    v^{\alpha',\nu}
                        \tau_\nu,
    \end{aligned}
    \end{eqnarray}
    where $\nu$ is the shorthand of $\boldsymbol{q}\lambda$, $\boldsymbol{q}$ the wave vector, $\lambda$ the index of phonon branch, $v^{\alpha,\nu}$ the phonon group velocity in direction $\alpha$,
    $C_\nu$ the phonon heat capacity, 
    and $\tau_{\nu}$ the phonon relaxation time, $N_0$ the number of unitcells in supercell, $\Omega$ the volume of unitcell.
    The phonon heat capacity is:
    \begin{eqnarray}
    \begin{aligned}
        \label{eq_phonon_heat_capacity}
        C_\nu = \frac{(\hbar\omega_v)^2 n_\nu(n_\nu+1)}{k_B T^2},
    \end{aligned}
    \end{eqnarray}
    where $n_\nu$ is Bose-Einstein Distribution:
    \begin{eqnarray}
    \begin{aligned}
        \label{eq_Bose-Einstein}
        n_\nu = \frac{1}{e^{\omega_\nu/k_B T}-1},
    \end{aligned}
    \end{eqnarray}
    $\omega_\nu$ is the phonon frequency.
    The relaxation time is inversely proportional to the sum of scattering matrix:
    \begin{eqnarray}
    \begin{aligned}
        \label{eq_relaxation_time}
        \frac{1}{\tau_\nu}
        =
        \bar{\Gamma}_\nu
        =
        \sum_{\nu^{ \prime} \nu^{ \prime \prime}}
            \left(\bar{\Gamma}_{\nu, \nu^{ \prime }}^{\nu^{ \prime\prime}}+\frac{\bar{\Gamma}^{\nu^{\prime \prime} ,\nu^{ \prime}}_{\nu}}{2}\right).
    \end{aligned}
    \end{eqnarray}
    The scattering matrix are:
    \begin{eqnarray}
    \begin{aligned}
        \label{eq_scattering_matrix}
        \bar{\Gamma}_{\nu, \nu^{ \prime }}^{\nu^{ \prime\prime}}
        =&\frac{2\pi}{\hbar}
            (n_{\nu'}-n_{\nu''})
                \left| 3!\cdot \tilde{\Phi}^{(3)}_S({\nu,\nu',-\nu''}) \right|^2
                    \\
                    &\times
                    \delta\left[-\hbar\omega_\nu-\hbar\omega_\nu'+\hbar\omega_\nu''\right],
        \\
        \bar{\Gamma}_{\nu}^{\nu^{ \prime}, \nu^{ \prime \prime}}
        =& \frac{2\pi}{\hbar}
            (1+n_{\nu'}+n_{\nu''})
                \left| 3!\cdot \tilde{\Phi}^{(3)}_S({\nu,-\nu',-\nu''}) \right|^2
                    \\
                    &\times
                    \delta\left[-\hbar\omega_\nu-\hbar\omega_\nu'+\hbar\omega_\nu''\right].
    \end{aligned}
    \end{eqnarray}
    $\tilde{\Phi}^{(3)}_S({\lambda, \lambda' ,\lambda''})$ is the ph-ph interaction vertex which can be calculate using following equation
    \begin{eqnarray}
    \begin{aligned}
    \label{eq_ph_ph_interaction_strength}   
        \tilde{\Phi}^{(3)}_S({\nu, \nu', \nu''}) = & \frac{\hbar^{3/2}}{3! \sqrt{8 N_0}}  
        \sum_{\substack{
          \rho \rho' \rho''\\
          \alpha \alpha' \alpha''
        }}
        e^{\rho \alpha}  \phantom{}_{\nu} 
        e^{\rho' \alpha'} \phantom{}_{\nu'} 
        e^{\rho'' \alpha''}  \phantom{}_{\nu''}
        \\
        &
        \sqrt{\frac{1}{m_{\rho} \omega_{\nu}}}  
        \sqrt{\frac{1}{m_{\rho'} \omega_{\nu'}}}  
        \sqrt{\frac{1}{m_{\rho''}\omega_{\nu''}}}
        \\
        &
        \sum_{l' l''} \Phi^{(3)}_{\rho \alpha, \rho' \alpha',' \rho''\alpha''} (0, l', l'') 
        \textrm{} 
        \\
        &
        \text{e}^{i\boldsymbol{q} \cdot \boldsymbol{r}^{0\rho}} 
        \text{e}^{i\boldsymbol{q}' \cdot \boldsymbol{r}^{l' \rho'}}
        \text{e}^{i\boldsymbol{q}'' \cdot \boldsymbol{r}^{l'' \rho''}},
    \end{aligned}
    \end{eqnarray}
    where $e^{\kappa\alpha}{}_\nu$ is the eigenvector, $\boldsymbol{\Phi}^{(3)}$ the 3$^{rd}$ order interatomic force constants, $\boldsymbol{r}^{l\rho}$ is the position of $\rho$-th atom  in $l$-th unitcell. In the main text, we use $a$ as shorthand of $l\rho$.
    The phonon properties above are determined by temperature dependent atom positions $\boldsymbol{r}^{l\rho}$ and 2nd and 3rd interatomic force constants $\boldsymbol{\Phi}^{(3)}$ and $\boldsymbol{\Phi}^{(3)}$. 
    
    When we calculate the scaled thermal conductivity $\overline{\kappa}$ at temperature $T$, we just replace the Bose-Einstein Distribution in equation Eq.~\eqref{eq_phonon_heat_capacity} and Eq.~\eqref{eq_scattering_matrix} with Bose-Einstein Distribution at 300 K:
    \begin{eqnarray}
    \begin{aligned}
        n_\nu \rightarrow \frac{1}{\exp[\omega_\nu/(k_B \cdot 300\ \mathrm{K})]-1}.
    \end{aligned}
    \end{eqnarray}

\section{The Scaled Thermal Conductivity Using Constant $ph-ph$ Interaction Strength}\label{const_PP}
\begin{figure}[htbp]
    \includegraphics{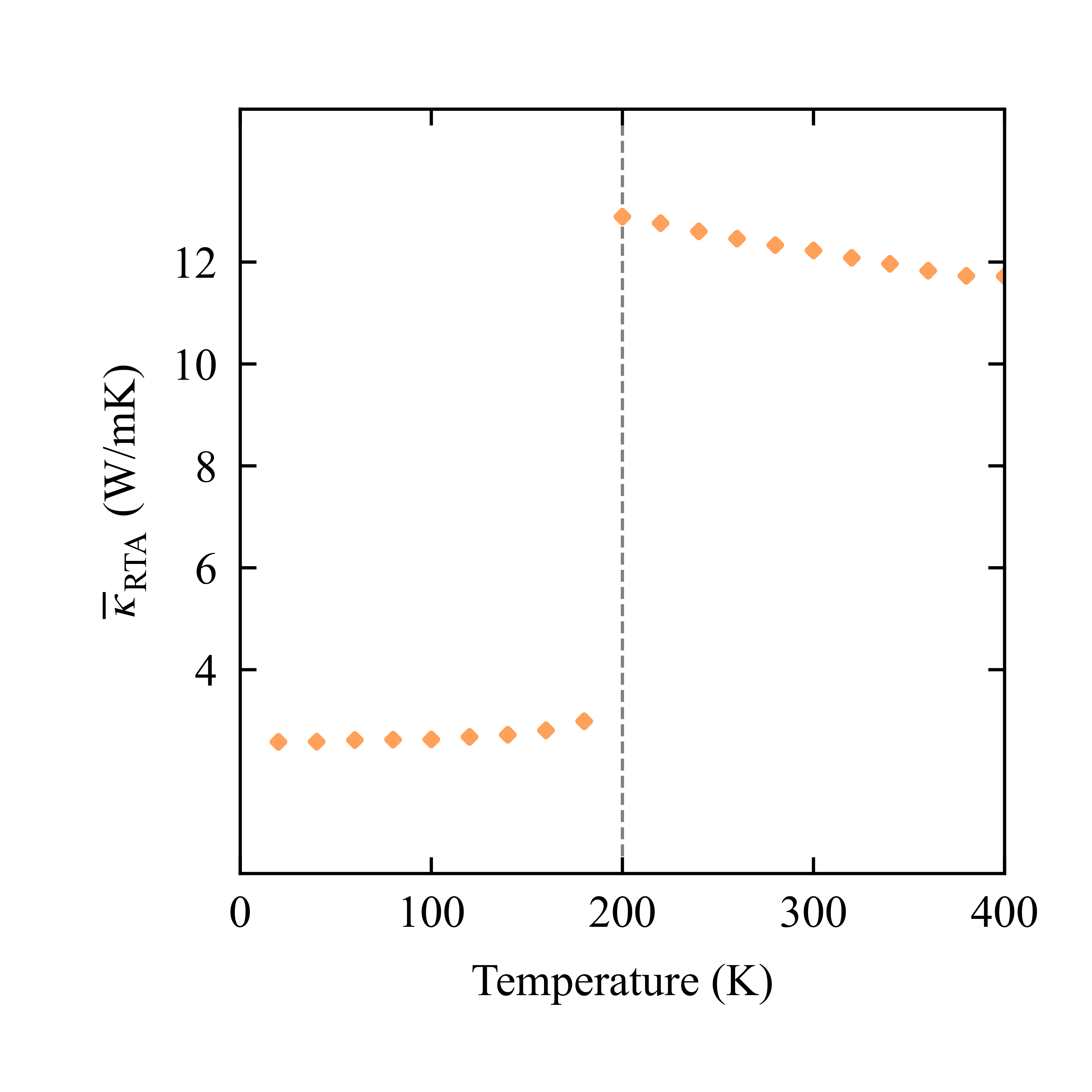}
    \caption{\label{fig_RTA_const_pp} 
    The scaled thermal conductivity using constant \textit{ph-ph} interaction strength $10^{-9}\textrm{eV}^2$.
    }
    \end{figure}

    Fig.~\ref{fig_RTA_const_pp} shows the scaled thermal conductivity using constant \textit{ph-ph} interaction strength $10^{-9}\textrm{eV}^2$.

\section{Soft Mode Frequencies and RTA Thermal Conductivity Using IFCs expanded around $\boldsymbol{r}_0$}\label{sec_only_loops}
    \begin{figure}[htbp]
    \includegraphics{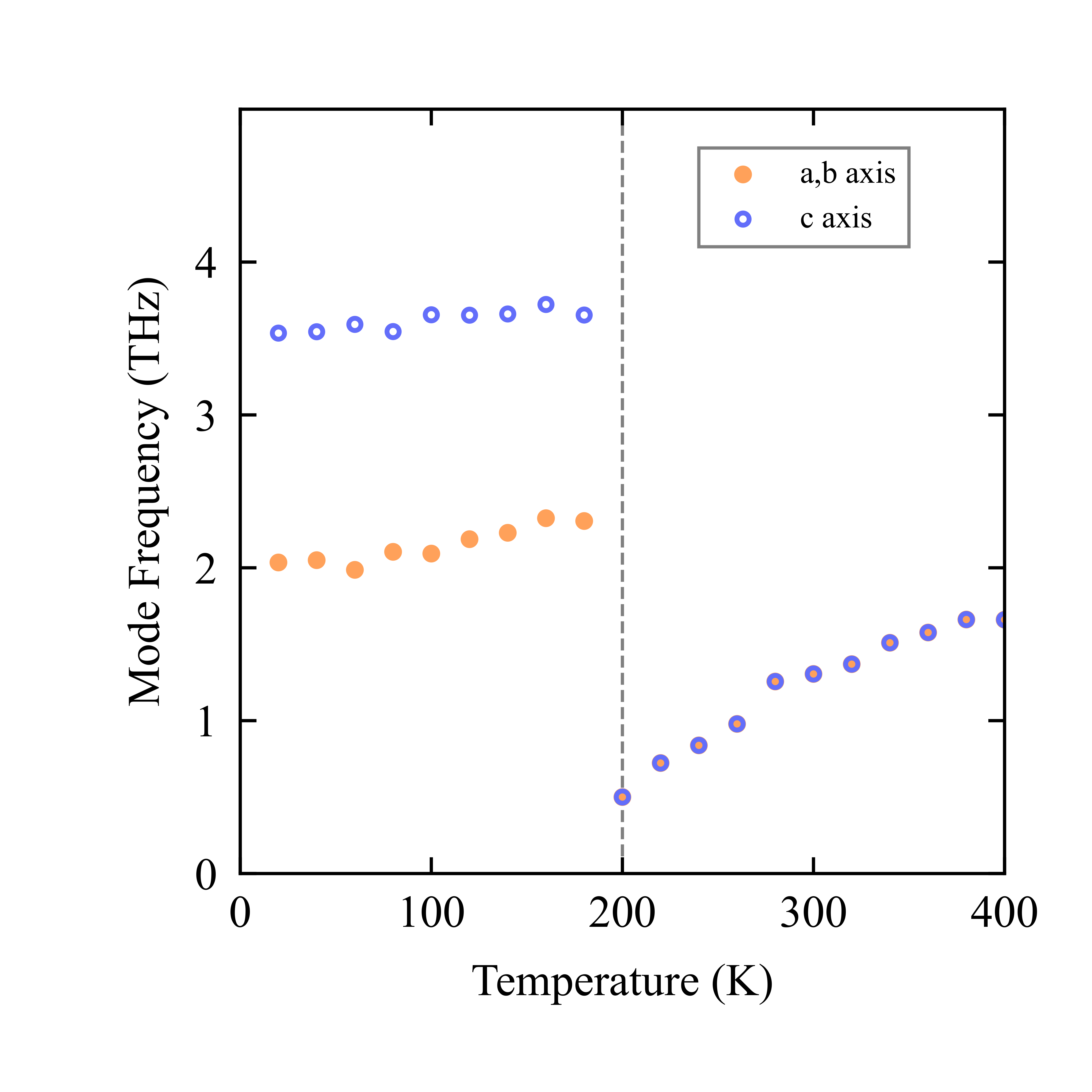}
    \caption{\label{fig_soft_mode_frequency_wo_r_T}
    The soft mode frequency using SCAILD, where the IFCs are expanded around $\boldsymbol{r}_0$ 
    }
    \end{figure}
    \begin{figure}[htbp]
    \includegraphics{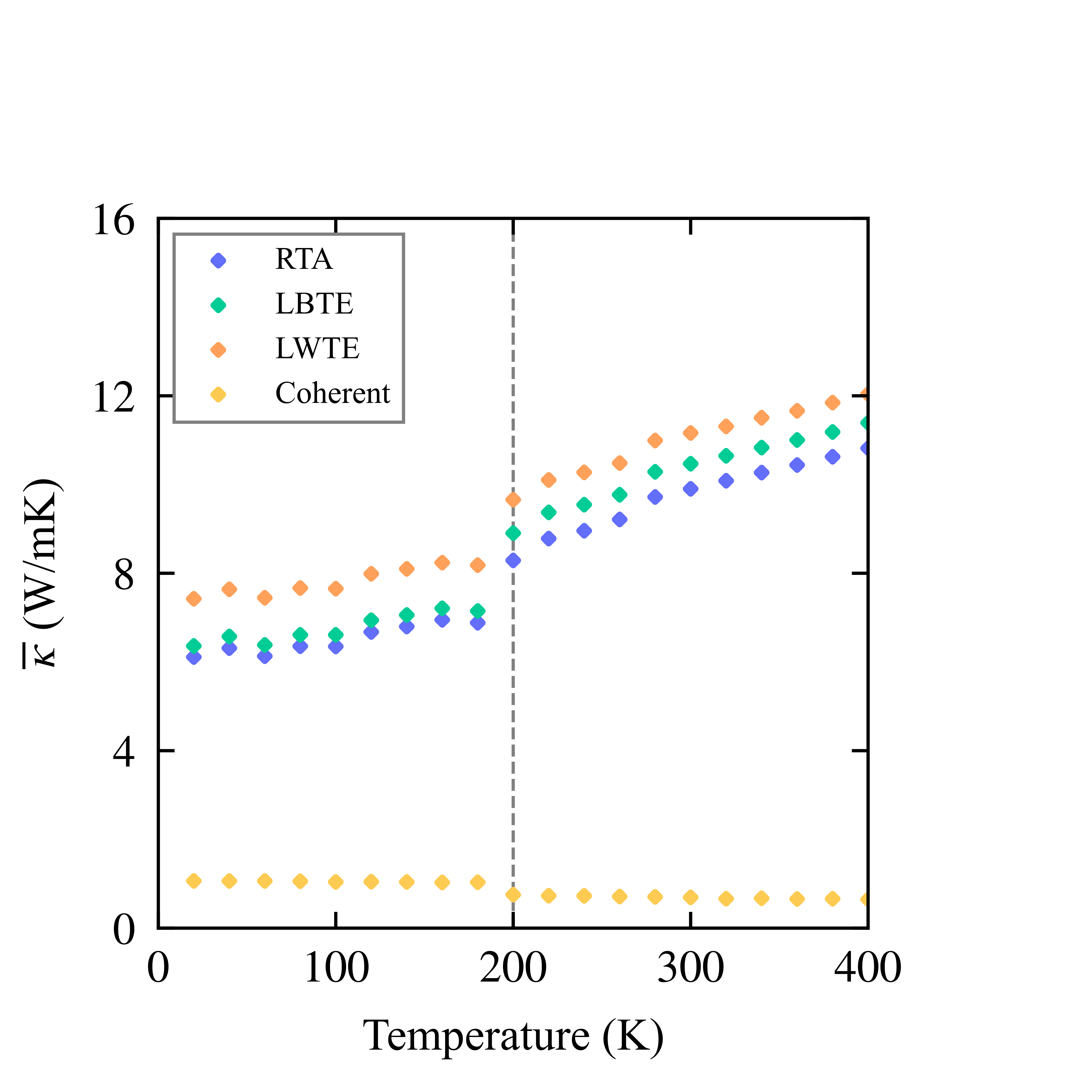}
    \caption{\label{fig_LWTE_results_wo_r_T} 
    The scaled thermal conductivity results where the IFCs are expanded around $\boldsymbol{r}_0$.
    }
    \end{figure}

    When the IFCs are expanded around $\boldsymbol{r}_0$, there will be a sharp change of $\tilde{\Phi}^{(2)}$ at the phase transition point, as shown in Fig.~\ref{main-fig_fc2_vs_T}.
    As a result, there will be also a sharp change in soft mode frequencies and thermal conductivity, as shown in Fig.~\ref{fig_soft_mode_frequency_wo_r_T} and Fig.~\ref{fig_LWTE_results_wo_r_T} which is not consistent with experimental results.

\section{Expressions of tadpole and loop self-energy terms}
The lowest order tadpole and loop self-energies are:
\begin{eqnarray}
    \begin{aligned}
        \label{eq_tadpole_loop_expression} 
        {\Sigma}^{\text{loop,0}}_{\nu}
          =&
          -\sum_{\nu_{1}} \tilde{\Phi}^{(4)}_0\left(\nu ;-\nu ; \nu_{1} ;-\nu_{1}\right) \frac{2 n_{\nu_1}+1}{2} \\
        {\Sigma}^{\text{tadpole,0}}_{\nu}
          =&
           -\frac{1}{\hbar} 
             \sum_{\lambda_{1}, \nu_{2}} 
               \frac{2 n_{\nu_2}+1}{\omega_{\mathbf{0}\lambda_{1}} }
                \\
          & \times 
          \tilde{\Phi}^{(3)}_0 \left(-\nu ; \nu ; \mathbf{0} \lambda_{1}\right) 
               \tilde{\Phi}^{(3)}_0 \left(\mathbf{0} \lambda_{1} ; \nu_{2} ;-\nu_{2}\right),\\
    \end{aligned}
\end{eqnarray}
where $\tilde{\boldsymbol{\Phi}}^{(n)}_0$ is the bare n\textsuperscript{th}-order \textit{ph-ph} interaction vertex.
% \nocite{*}
% \clearpage
\bibliographystyle{apsrev4-2}
\bibliography{main}% Produces the bibliography via BibTeX.